\documentclass[12pt]{iopart}
\usepackage{iopams}
\usepackage{graphicx}
\begin{document}

\title[Nonclassical tests]{Independent nonclassical tests
for states and measurements in the same experiment}

\author{Alfredo Luis$^1$ and \'{A}ngel Rivas$^2$}

\address{$^1$ Departamento de \'{O}ptica, Facultad de Ciencias
F\'{\i}sicas, Universidad Complutense, 28040 Madrid, Spain}
\address{$^2$ Institut f\"{u}r Theoretische Physik, Universität Ulm,
Ulm D-89069, Germany}
\ead{alluis@fis.ucm.es}

\begin{abstract}
We show that one single experiment can test simultaneously and
independently both the nonclassicality of states and measurements
by the violation or fulfillment of classical bounds on the statistics.
Nonideal measurements affected by imperfections can be characterized
by two bounds depending on whether we test the ideal measurement or
the real one.
\end{abstract}

%\pacs{03.65.Ca, 03.65.Ta, 42.50.Dv, 42.50.Ar}

\noindent{\it Keywords}: Nonclassical states, quantum optics

%\submitto{\PS}

\maketitle

\section{Introduction}

Within standard quantum theory, quantum states play two dissimilar
but complementary roles: (i) they express the state of the system,
represented by a density matrix $\rho$, and (ii) they determine
the statistics of measurements, typically by projection of the
system state on the eigenstates of the measured observable. More
precisely, any observable event is represented by a nonnegative
Hermitian operator $\Delta$ (maybe part of a larger positive
operator-valued measure) that determines the event probability
as $p = \tr ( \rho  \Delta )$, where $\rho$ is the state
of the system. In many relevant situations $\Delta$ is proportional
to a suitable state, such as photon-number and quadrature
measurements in quantum optics. Positive operators playing the
role of $\Delta$ can be turned into the role $\rho$ as shown in
\cite{L}.

While referring to nonclassical states is quite common, not so
many effort has been devoted to nonclassical measurements
\cite{M1,M2,nos,Amri}. A customary criterion of nonclassicality
for states $\rho$ is the failure of the Glauber-Sudarshan $P$
function to exhibit all the properties of a classical probability
density \cite{bo}. This occurs when $P$ takes negative values,
or when it fails to be a proper function becoming more singular
than the delta function. Accordingly, we may say that the event
represented by $\Delta$ in Hilbert space is nonclassical when
its $P$ phase-space representative takes negative values or is
more singular than the delta function.

Although the nonclassicality of states and measurements are
different things, both may be tested simultaneously within
one single experiment in terms of its statistics $p$. This
possibility is addressed in this work by means of a simple
example: nonefficient single-photon detection in photon-added
thermal states \cite{AT}. A key point of this example is
feasibility, since these states have been already generated
in experiment \cite{exp}, and it explicitly includes typical
imperfections such as losses and thermalization.

As nonclassicality criteria we will consider the simple and
robust practical tests recently introduced where nonclassicality
is revealed by breaking classical bounds on probabilities
satisfied by all classical states and measurements \cite{nos,MH}
(see \cite{AM} for other nonclassicality criteria). The main
features of these tests are recalled in section 2.

\section{Classical bounds on probabilities}

For definiteness let us focus on a single mode of the electromagnetic
field with complex-amplitude operator $a$. To derive the nonclassical
tests we will use the $P$ and $Q$ phase-space representatives associated
to any operator $A$
\begin{equation}
A = \int \rmd^2 \alpha P_A (\alpha ) | \alpha \rangle \langle
\alpha |, \quad
Q_A (\alpha ) = \frac{1}{\pi} \langle \alpha | A | \alpha \rangle ,
\end{equation}
where $| \alpha \rangle$ are coherent states, $a |\alpha \rangle =
\alpha | \alpha \rangle$. They are suitably normalized
\begin{equation}
\label{nor}
\int \rmd^2 \alpha P_A (\alpha) = \int \rmd^2 \alpha Q_A (\alpha)
= \tr A,
\end{equation}
with $\rmd^2 \alpha = \rmd x \rmd y $, where $x$, $y$ are the real
and imaginary parts of $\alpha = x+ \rmi y$.

Exploiting the $\rho \leftrightarrow \Delta$ symmetry, the same
probability $p = \tr ( \rho \Delta )$ can be expressed by two
equivalent formulas
\begin{equation}
\label{pm}
p = \pi
\int \rmd^2 \alpha P_\rho  (\alpha) Q_\Delta (\alpha) = \pi \int
\rmd^2 \alpha P_\Delta (\alpha) Q_\rho (\alpha) .
\end{equation}
By using the first equality we are able to derive bounds
sensitive to the nonclassicality of the state $\rho$, while
the second equality leads to bounds sensitive to the
nonclassicality of the measurement $\Delta$. Note that the
$Q$ function is always a positive and well behaved.

\subsection{Nonclassical test for states}

For classical states, i. e., for ordinary nonnegative functions
$P_\rho (\alpha) \geq 0$, we get
\begin{equation}
\label{pP}
P_\rho (\alpha) Q_\Delta (\alpha) \leq P_\rho  (\alpha) Q_{\Delta,
\mathrm{max}},
\end{equation}
where $Q_{\Delta,\mathrm{max}}$ is the maximum of $Q_\Delta (\alpha)$
when $\alpha$ is varied. Applying this to the first equality in
(\ref{pm}), and taking into account (\ref{nor}), we get the
following upper bound $S$ for $p$
\begin{equation}
\label{trQP}
p \leq S = \pi Q_{\Delta ,\mathrm{max}} ,
\end{equation}
that holds for every $P_\rho (\alpha)$ compatible with classical
physics. If this condition is violated it means that (\ref{pP}) is
false and the state is not classical.

\subsection{Nonclassical test for measurements}

For classical measurements, i. e., for ordinary nonnegative
function $P_\Delta (\alpha) \geq 0$ it holds that
\begin{equation}
\label{pPm}
P_\Delta (\alpha) Q_\rho (\alpha) \leq P_\Delta (\alpha)
Q_{\rho ,\mathrm{max}},
\end{equation}
where $Q_{\rho, \mathrm{max}}$ is the maximum of $Q_\rho (\alpha)$
when $\alpha$ is varied. Applying this to the second equality in
(\ref{pm}) we get the following upper bound $M$ for $p$, provided
that $\tr \Delta$ is finite,
\begin{equation}
\label{trPQ}
p \leq M = \pi Q_{\rho, \mathrm{max}} \tr \Delta.
\end{equation}
Equation (\ref{trPQ}) can be violated if $P_\Delta (\alpha)$
fails to be positive or when it becomes a generalized function
(this is a nonclassical measurement) since in both cases
(\ref{pPm}) fails to be true.

\section{Inefficient photon detection on photon-added thermal
states}

The same probability $p$ may serve to test both the nonclassicality
of $\rho$ and $\Delta$. Let us demonstrate this by applying the
above formalism to nonefficient single-photon detection in
photon-added thermal states.

\subsection{Single-photon-added thermal states}

The single-photon-added thermal states read, in the photon-number
basis, \cite{AT,exp}
\begin{equation}
\label{pats}
\rho = (1 - \xi) a^\dagger \rho_{\mathrm{tc}} a = (1 - \xi)^2
\sum_{n=1}^\infty \xi^{n-1} n | n \rangle
\langle n | ,
\end{equation}
where $\rho_{\mathrm{tc}}$ is a thermal chaotic state
\begin{equation}
\label{tc}
\rho_{\mathrm{tc}} = \left ( 1 - \xi \right ) \sum_{n=0}^\infty
\xi^n | n \rangle \langle n | ,
\end{equation}
with mean number of photons
\begin{equation}
\label{ntc}
\bar{n} =\frac{\xi}{1-\xi} .
\end{equation}
The $P$ representative of $\rho$ is well-behaved but nonpositive
\begin{equation}
\label{Pr}
P_\rho (\alpha) = \frac{1}{\pi \bar{n}^3} \left [ \left (
\bar{n} + 1 \right ) | \alpha |^2 - \bar{n}
\right ] \exp \left ( - \frac{| \alpha |^2}{\bar{n}} \right ) ,
\end{equation}
while the $Q$ function is
\begin{equation}
Q_\rho (\alpha) = \frac{| \alpha |^2}{\pi (\bar{n}+1)^2}
\exp \left ( - \frac{| \alpha |^2}{\bar{n} +1} \right ) ,
\end{equation}
so that its maximum occurs for $|\alpha |^2 = \bar{n}+1$
being
\begin{equation}
\label{Qrm}
Q_{\rho, \mathrm{max}} = \frac{1}{\pi \rme (\bar{n}+1)} .
\end{equation}
These states present three relevant features for our purposes:

(i) Their nonclassical behavior is independent of other typical
nonclassical features, since there is no quadrature squeezing,
they present super-Poissonian photon-number statistics for all
$\bar{n} > 1/\sqrt{2}$, and have no oscillatory statistics
\cite{nos}.

(ii) They can be generated experimentally \cite{exp}.

(iii) Their definition embodies a typical source of practical
imperfection such as thermalization.

\subsection{Ideal single-photon detection}

For ideal single-photon detection we have, in the photon-number
basis, $\Delta = | 1 \rangle \langle 1 |$ with $\tr \Delta =1$.
The $P$ representative is nonclassical being more singular than the
delta function
\begin{equation}
P_\Delta (\alpha) = \left ( 1+ \frac{\partial^2}{\partial \alpha
\partial \alpha^\ast} \right )  \delta^{(2)} (\alpha) .
\end{equation}
The $Q$ function is
\begin{equation}
\label{QD}
Q_\Delta (\alpha) = \frac{|\alpha|^2}{\pi} \exp(-|\alpha|^2) ,
\end{equation}
and the maximum occurs at $|\alpha|=1$
\begin{equation}
Q_{\Delta,\mathrm{max}} = \frac{1}{\rme \pi}.
\end{equation}
If the measured state is classical, the single-photon
probability $p$ is bounded by \cite{nos,MH}
\begin{equation}
\label{cpr}
p \leq S = \frac{1}{\rme} .
\end{equation}

\subsection{Inefficient single-photon detection}

In figure 1 we illustrate the case of inefficient single-photon
detection. A detector with quantum efficiency $\eta$ can be
modeled by a beam splitter of amplitude-transmission coefficient
$t = \sqrt{\eta}$, mixing the input state $\rho$ with vacuum,
placed before an ideal detector $\Delta$ with $\eta=1$ \cite{qeta}.
After this model two different routes can be followed:

\begin{figure}
\begin{center}
\includegraphics{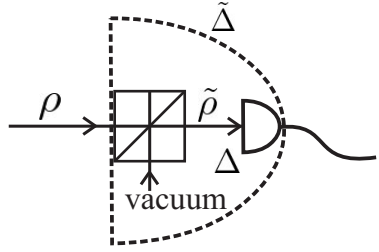}
\end{center}
\caption{Illustration of inefficient single-photon detection.}
\end{figure}

(i) We can test the underlying ideal detection $\Delta$ regarding
inefficiency as a handicap of practical origin. This is to say that
we have ideal detection on the state $\tilde{\rho}$ after the beam
splitter that carries the effect of inefficiency, so that the
probability is $p= \tr ( \tilde{\rho} \Delta )$. Since the
transformation of coherent states through lossless beam splitters
is $| \alpha \rangle \rightarrow | \sqrt{\eta} \alpha \rangle$ the
state $\tilde{\rho}$ is
\begin{equation}
\label{tiro}
\tilde{\rho} = \int \rmd^2 \alpha P_{\rho} (\alpha)
| \sqrt{\eta} \alpha \rangle \langle  \sqrt{\eta} \alpha | ,
\end{equation}
where $P_{\rho} (\alpha)$ is in (\ref{Pr}). From this
expression we get by direct computation the $Q$ function
of $\tilde{\rho}$
\begin{equation}
Q_{\tilde{\rho}} (\alpha) = \frac{1}{\pi} \left [
\frac{(\bar{n}+1) \eta | \alpha |^2}{(\eta \bar{n} + 1)^3}
+ \frac{1-\eta}{(\eta \bar{n} + 1)^2} \right ] \exp
\left ( - \frac{| \alpha |^2}{\eta \bar{n} +1} \right ) .
\end{equation}
For $\eta \neq 0$ its maximum holds for
\begin{equation}
| \alpha |^2 = 1+ \eta \bar{n} -
\frac{(1-\eta)(1+ \eta \bar{n})}{\eta (\bar{n}+1)} ,
\end{equation}
leading to
\begin{equation}
\label{Qtrm}
Q_{\tilde{\rho},\mathrm{max}} =
\frac{\eta (\bar{n}+1)}{\pi (\eta \bar{n} + 1)^2}
\exp \left [ - \frac{\eta \bar{n} +2\eta -1}{\eta
(\bar{n} +1)} \right ] .
\end{equation}

(ii) Alternatively, we can test the real measurement embodying
the inefficiency as part of the measuring apparatus. The real
measurement is represented by an Hermitian nonnegative operator
$\tilde{\Delta}$ to be determined such that the probability
can be expressed as $p  = \tr ( \rho \tilde{\Delta} )$. From
(\ref{pm}), (\ref{tiro}) and the equality $p= \tr ( \tilde{\rho}
\Delta )= \tr ( \rho \tilde{\Delta} )$ we get
\begin{equation}
\label{QtD}
Q_{\tilde{\Delta}} (\alpha) = Q_{\Delta} ( \sqrt{\eta} \alpha)
= \frac{\eta |\alpha|^2}{\pi} \exp(- \eta |\alpha|^2) ,
\end{equation}
so that
\begin{equation}
\label{iq}
Q_{\tilde{\Delta}, \mathrm{max}} =
Q_{\Delta, \mathrm{max}} = \frac{1}{\pi \rme} .
\end{equation}
From (\ref{nor}), (\ref{QtD}) and $\tr \Delta =1$ we readily get
\begin{equation}
\label{tr}
\tr \tilde{\Delta} = \frac{1}{\eta} .
\end{equation}
Moreover, by expressing the exponential in (\ref{QtD}) as $\exp(
- \eta |\alpha|^2) = \exp[(1- \eta )|\alpha|^2] \exp(- |\alpha|^2)$
and expanding the first exponential in power series we get
\begin{equation}
Q_{\tilde{\Delta}} (\alpha) = \frac{\eta }{\pi} \sum_{n=0}^\infty
(n+1)(1-\eta )^n \frac{|\alpha|^{2(n+1)}}{(n+1)!} \exp(- |\alpha|^2),
\end{equation}
that readily provides the expression of $\tilde{\Delta}$ in the
photon-number basis
\begin{equation}
\label{tD}
\tilde{\Delta} = \eta \sum_{n=0}^\infty (n+1) (1-\eta )^n
|n+1 \rangle \langle n+1 | = \eta (1-\eta )^{a^\dagger a -1}
a^\dagger a .
\end{equation}

The probability $p$ is independent of the interpretations (i) and
(ii), being [from (\ref{pats}) and (\ref{tD}) for example]
\begin{equation}
\label{pt12}
p = \tr ( \tilde{\rho} \Delta ) =
\tr ( \rho \tilde{\Delta} ) =  \eta \frac{1 + 2
\bar{n} - \eta \bar{n}}{ \left ( 1+ \eta
\bar{n} \right )^3} .
\end{equation}
The interpretations also do not affect the classical upper
bound for states, from (\ref{trQP}) and (\ref{iq}),
\begin{equation}
\label{cpr2}
p \leq S = \pi Q_{\Delta ,\mathrm{max}} = \frac{1}{\rme} .
\end{equation}

The interpretations (i)/(ii) affect the classical upper bound
for measurements. Since we are actually considering two different
measurements, the ideal $\Delta$ and the real $\tilde{\Delta}$,
we get from (\ref{Qrm}), (\ref{Qtrm}), and (\ref{tr}) two
different classical bounds $M_\Delta$ and $M_{\tilde{\Delta}}$ with
\begin{equation}
\label{trPQ1}
p \leq M_\Delta = \pi Q_{\tilde{\rho},\mathrm{max}}
\tr \Delta =  \frac{\eta (\bar{n}+1)}{(\eta \bar{n}
+ 1)^2} \exp \left [ - \frac{\eta \bar{n} +2 \eta -1}{\eta
(\bar{n} +1)} \right ] ,
\end{equation}
and
\begin{equation}
\label{trPQ2}
p \leq M_{\tilde{\Delta}} = \pi Q_{\rho,\mathrm{max}}
\tr \tilde{\Delta} =  \frac{1}{\rme \eta (\bar{n}+1)} .
\end{equation}

In figure 2 we have plotted $p$, $M_{\tilde{\Delta}}$,
$M_\Delta$, and $S$ as functions of $\bar{n}$ for $\eta
= 0.4$ and $0.9$, while in figure 3 they are plotted as
functions of $\eta$ for $\bar{n} = 0.2$ and $0.7$.
Several conclusions can be derived from these plots:

(a) $M_{\tilde{\Delta}}$ is always above $M_\Delta$, so
that it is more difficult to prove the nonclassicality
of the real $\tilde{\Delta}$ than of the ideal $\Delta$.
This fits with the general idea that inefficiencies degrade
quantum properties.

(b) When $\eta$ increases $M_{\tilde{\Delta}}$ and $M_\Delta$
become closer. This may be expected since when $\eta
\rightarrow 1$ we get $\tilde{\Delta} \rightarrow \Delta$.

(c) In general $p$ decreases when $\bar{n}$ increases,
enforcing the fulfillment of the classical bounds. This agrees
with common understanding of the effect of thermalization.
An exception occurs when $\eta$ is rather low, since for small
$\eta$ thermal photons may increase the probability of photon
detection, as illustrated in figure 2(a).

(d) For low values of $\bar{n}$ increasing $\eta$ favors the
violation of the classical bounds by increasing $p$ and
decreasing $M_{\tilde{\Delta}}$ and $M_\Delta$, as illustrated
in figure 3(a). On the other hand, for larger $\bar{n}$ we get
that larger $\eta$ increases the probability of detecting
more than a single photon, decreasing $p$ as illustrated in
figure 3(b).

(e) For large $\eta$ and small $\bar{n}$ it is possible to have
$p > M_{\tilde{\Delta}}, M_\Delta, S$ simultaneously,
so that one and the same measurement can reveal at the same time
the nonclassical character of the $\rho$, $\Delta$, and
$\tilde{\Delta}$ as illustrated in figures 2(b) and 3(a).

\begin{figure}
\begin{center}
\includegraphics{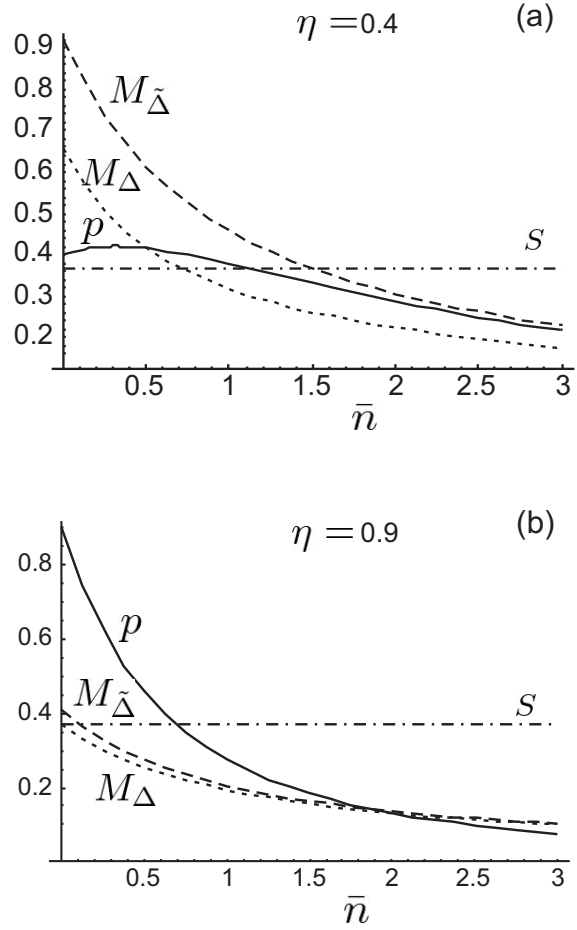}
\end{center}
\caption{Plots of $p$ (solid), $M_{\tilde{\Delta}}$ (dashed),
$M_\Delta$ (dotted), and $S$ (dash-dotted), as functions of
$\bar{n}$ for fixed $\eta = 0.4$ (a) and $\eta=0.9$ (b).}
\end{figure}

\begin{figure}
\begin{center}
\includegraphics{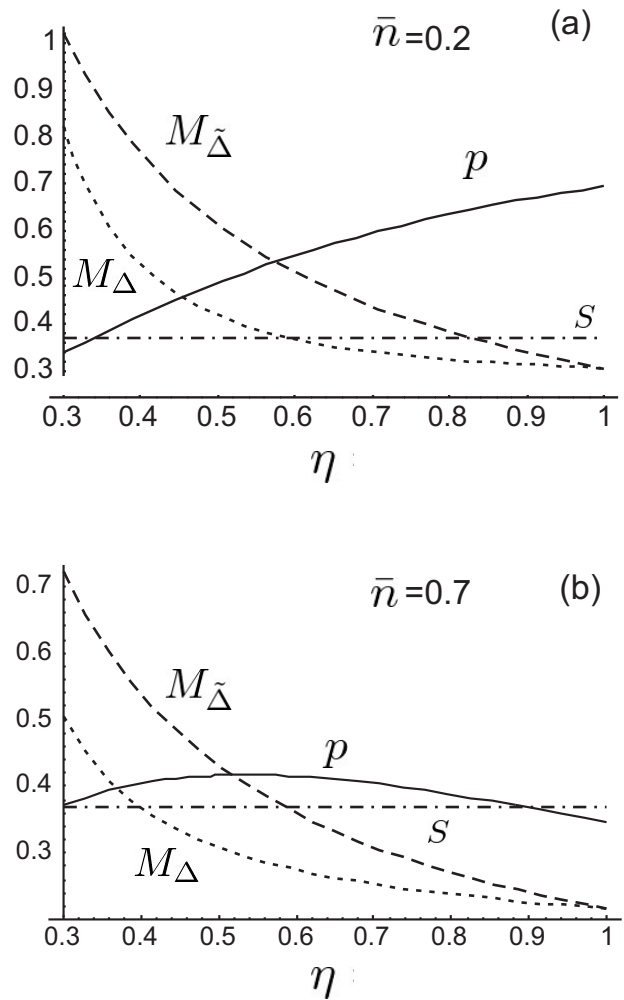}
\end{center}
\caption{Plots of $p$ (solid), $M_{\tilde{\Delta}}$ (dashed),
$M_\Delta$ (dotted), and $S$ (dash-dotted), as functions of
$\eta$ for fixed $\bar{n}= 0.2$ (a) and $\bar{n}=0.7$ (b).}
\end{figure}

\section{Conclusions}

We have shown that the same experiment can test simultaneously and
independently both the nonclassicality of states and measurements.
This is because the nonclassicality of states and measurements
manifests via the violation of different and independent bounds
to the same statistics. We have shown that practical imperfections
produce the existence of two bounds testing the ideal and real
measurements.

\ack

A. R. acknowledges financial support from the EU Integrated
Project QESSENCE and the STREP action CORNER. A. L. acknowledges
support from project No. FIS2008-01267 of the Spanish Direcci\'{o}n
General de Investigaci\'{o}n del Ministerio de Ciencia e
Innovaci\'{o}n, and by project QUITEMAD S2009-ESP-1594 of the
Consejer\'{\i}a de Educaci\'{o}n de la Comunidad de Madrid.

\end{document}